%% file: main.tex
\documentclass{egpubl}
\usepackage{eurovis2026}

\EuroVisShort  %

\usepackage[T1]{fontenc}
\usepackage{dfadobe}

\usepackage{cite}  %

\usepackage{booktabs}
\usepackage{subcaption}
\usepackage{xurl}

\BibtexOrBiblatex
\electronicVersion
\PrintedOrElectronic
\ifpdf \usepackage[pdftex]{graphicx} \pdfcompresslevel=9
\else \usepackage[dvips]{graphicx} \fi

\usepackage{egweblnk}

\usepackage{cleveref}

\providecommand{\alg}[1]{\textsc{#1}}

\newcommand{\methodGau}{\alg{Gaussian}}
\newcommand{\methodUni}{\alg{Uniform-Subsample}}
\newcommand{\methodDP}{\alg{Douglas-Peucker}}
\newcommand{\methodTDA}{\alg{Topolines}}
\newcommand{\methodASAP}{\alg{ASAP}}
\newcommand{\methodPAA}{\alg{PAA}}

\Crefformat{figure}{#2Fig.~#1#3}
\Crefmultiformat{figure}{Figs.~#2#1#3}{ and~#2#1#3}{, #2#1#3}{ and~#2#1#3}
\Crefformat{section}{#2Sect.~#1#3}
\Crefmultiformat{section}{Sect.~#2#1#3}{ and~#2#1#3}{, #2#1#3}{ and~#2#1#3}

\title{Beyond One-Size-Fits-All: User Strategies for Simplification Technique and Level Selection in Responsive Line Charts}

\author[Proma \& Rosen]
{\parbox{\textwidth}{\centering
Rifat Ara Proma\orcid{0009-0000-2557-3188} and Paul Rosen\orcid{0000-0002-0873-9518}
}
\\
{\parbox{\textwidth}{\centering
Scientific Computing and Imaging Institute, University of Utah, Salt Lake City, Utah, USA
}}
}

\begin{document}
\teaser{
    \centering
    \begin{minipage}[t]{\linewidth}
        \centering
        \includegraphics[width=0.88\linewidth]{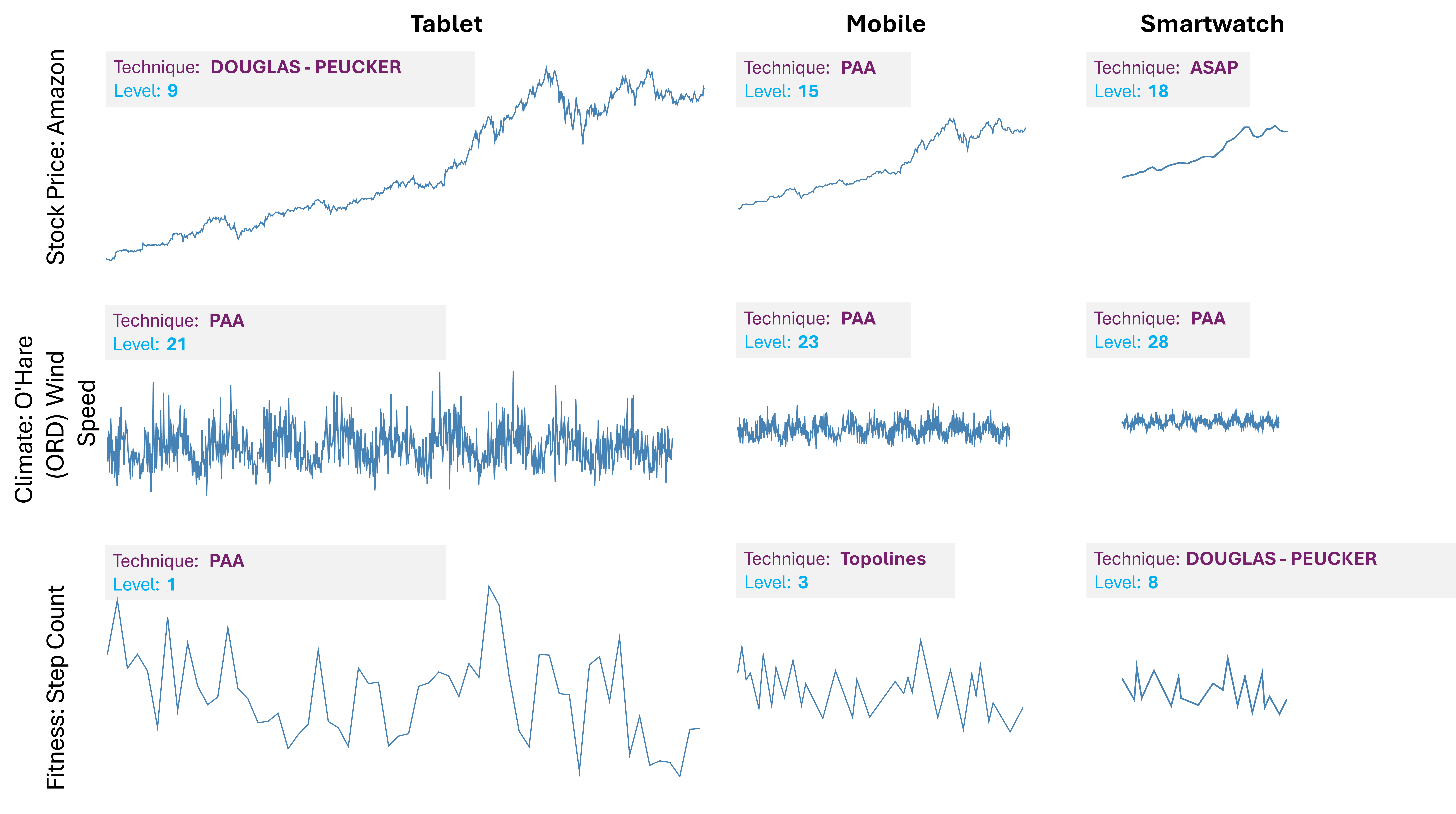}

        \caption{
            The figure shows the final simplified charts generated by different participants under the C2 condition. Participants selected different simplification techniques and levels depending on the dataset and device size.
        }
        
        \label{fig:teaser}
    \end{minipage}
  
}

\maketitle
\begin{abstract}
Simplifying line charts for responsive displays typically applies a single algorithm uniformly across devices, despite the availability of multiple techniques that preserve different signal characteristics (e.g., peaks, trends, periodicity). We investigate whether users benefit from algorithmic choice when adapting charts across screen sizes. In a within-subjects study (N=30), participants simplified nine datasets under three conditions: single pre-assigned technique (C1), multiple techniques (C2), and multiple techniques with manual point selection (C3), each with control over simplification level. We found that users adapted technique selections across datasets rather than devices, leveraging dataset-level strategies rather than per-device optimization. Additionally, interaction complexity did not always increase engagement uniformly, suggesting that responsive simplification tools should balance algorithmic flexibility with progressive disclosure and strong defaults. Supplemental materials are available at \url{https://osf.io/yjp76/?view_only=b77b5e97f0cc4f689fbf48ad0d965af3}.


\end{abstract}  

\input{sec.introduction.tex}

\input{sec.prior_work}
\input{sec.simplification_strategy}
\input{sec.study_design}

\input{sec.results}

\input{sec.discussion}
\input{sec.acknowledgement}

\bibliographystyle{eg-alpha-doi} 
\bibliography{main} 
\end{document}

%% file: sec.introduction.tex
\section{Introduction}
The rapid expansion of mobile device usage has transformed how people access and interpret data visualizations~\cite{schottler2024practices, kim2021design}. Charts now appear across dashboards, analytical tools, and mobile interfaces, communicating diverse data such as climate patterns, health metrics, and stock prices. Often, these are time-series data which are commonly represented as line charts because they effectively convey trends, extrema, and anomalies~\cite{few2008line, aigner2007visualizing}. However, rendering all data points on small screens often produces visual clutter, requiring simplification through aggregation, filtering, or subsampling~\cite{schottler2024constraint, setlur2021semantic, elmqvist2009hierarchical}.

Responsive visualization research has explored annotation removal, encoding changes, layout adaptation~\cite{kim2021design, hoffswell2020techniques}, and geometric resizing techniques~\cite{wu2012visizer}. Yet simplification of the underlying data is typically handled by applying a single algorithm uniformly across device sizes~\cite{setlur2021semantic, elmqvist2009hierarchical}. Despite the existence of many simplification techniques with distinct preservation behaviors~\cite{rosen2020linesmooth, van2023data}, little is known about whether users benefit from selecting among them, or whether such flexibility introduces unnecessary complexity.

Prior work shows that simplification effectiveness depends on both algorithmic behavior and human factors~\cite{rosen2020linesmooth,proma2025visual}. Automated approaches to responsive visualization~\cite{setlur2021semantic, kim2021automated, wu2012visizer} may not always reflect designer intent, and discrepancies between design goals and viewer perception may occur~\cite{quadri2024you}. Without control over the simplification strategy, designers cannot ensure that key features are consistently preserved across device sizes.

We examined whether users prefer access to multiple simplification techniques and what level of control they value: technique selection, level adjustment, or manual point specification. We provided six algorithmically diverse techniques: \methodASAP{}~\cite{rong2017asap}, \methodDP{}~\cite{douglas1973algorithms}, \methodGau~\cite{kopparapu2011identifying}, \methodPAA~\cite{keogh2001dimensionality}, \methodTDA~\cite{suh2019topolines}, and \methodUni. In a study with 30 Prolific participants configuring nine datasets across three device sizes, we compared three conditions: single technique with level control (C1), multiple techniques with level control (C2), and C2 with additional manual point selection (C3).

We found that 56.7\% of participants preferred having multiple simplification techniques to choose from when creating charts for different device sizes. However, participants also preferred the efficiency of single simplification techniques with a level adjuster. Our contributions include: \textbf{(1)} empirical evidence that users leverage algorithmic choice for dataset-level adaptation rather than device-level re-optimization; \textbf{(2)} identification of adaptive behaviors across control dimensions, where restricted technique choice led to compensatory slider exploration, but added manual control (point selection) increased disengagement; and \textbf{(3)} demonstration that technique selection is highly individualized even within dataset categories, challenging one-size-fits-all approaches.

%% file: sec.prior_work.tex
\section{Related Work}
Responsive visualization research highlights that resizing introduces trade-offs between preserving message clarity and maintaining data density. Studies of communication-oriented graphics identify strategies such as rescaling, filtering, re-encoding, and element removal, while demonstrating that these changes can threaten legibility across devices~\cite{hoffswell2020techniques,kim2021design}. Subsequent work has formalized responsiveness through task-driven loss functions and mixed-initiative systems that balance density and communicative goals~\cite{kim2021automated,kim2023dupo}.

Another line of research frames responsive design as an optimization or constraint-solving problem. Perception-based resizing methods use clutter and saliency metrics to preserve important regions during deformation~\cite{wu2012visizer,li2018fast}, while broader surveys position such techniques as canonical retargeting problems in visualization~\cite{sun2023application}. More recent work replaces heuristic breakpoints with systematic constraint-based adaptation rules across screen sizes~\cite{schottler2024constraint}.

Informed by cartographic generalization, semantic resizing approaches prioritize important elements, such as extrema and endpoints, when space is reduced~\cite{setlur2021semantic}. Responsive thematic maps similarly employ element displacement, merging, or substitution to manage complexity~\cite{schottler2024practices,opach2019visual}. Complementarily, user-adaptive visualization research emphasizes personalization based on traits, goals, and cognitive states, though scalable implementations remain limited~\cite{yanez2025state}.

Together, these prior works advance layout adaptation, semantic filtering, and optimization frameworks for responsive visualization. However, these approaches primarily focus on layout transformation and element selection, treating underlying data transformation as secondary. Data simplification methods, such as subsampling, filtering, and aggregation, offer direct mechanisms for controlling density while preserving meaningful structures in line charts. Yet, responsive visualization research has not systematically examined the selection of simplification techniques as a design dimension. Our work addresses this gap by positioning data simplification as a primary mechanism for adaptive line chart design across screen sizes.

%% file: sec.simplification_strategy.tex
\section{Simplification Strategy}

We used nine real-world time-series datasets spanning three domains to ensure diversity in trend structure and variability. The datasets include stock prices (Apple, Amazon, Google)~\cite{yahoo_finance}, climate data (average wind speed from ATL, ORD, and SEA airports)~\cite{noaa}, and fitness metrics (burned kilocalories, miles covered, and step count from pedometer data)~\cite{pedometer}. These data types are commonly viewed across devices of varying sizes (e.g., financial dashboards, weather apps, and health trackers), making them representative contexts for responsive visualization. Each category contains three datasets with different signal characteristics (e.g., volatility, periodicity, and noise levels), enabling evaluation of simplification behavior across different temporal patterns.

We selected six simplification techniques to represent algorithmically diverse strategies grounded in prior literature and practice. \methodGau{} and \methodTDA{} were included based on strong performance in prior task-based evaluations~\cite{rosen2020linesmooth}. \methodDP{} was chosen due to its adoption in responsive chart systems~\cite{setlur2021semantic}, and Piecewise Aggregate Approximation (\methodPAA{}) for its use in hierarchical aggregation approaches~\cite{elmqvist2009hierarchical}. \methodASAP{} was included as an attention-aware smoothing method designed to reduce clutter while preserving salient patterns. Finally, \methodUni{} serves as a baseline sampling strategy without explicit feature prioritization. Together, these methods span smoothing-based, topology-aware, aggregation-based, and sampling-based strategies.

Because each technique relies on distinct parameters (e.g., deviation tolerance for \methodDP{}, persistence threshold for \methodTDA{}, bin count for \methodPAA{}, standard deviation for \methodGau{}), we normalize simplification strength in two stages. For each dataset of length $n$, we sample 100 parameter settings per technique using a logarithmic scale, generating 100 outputs per dataset–technique pair, where higher levels correspond to stronger simplification. We compute Pixel Approximate Entropy (PAE)~\cite{2018entropy} for each output to quantify simplification strength (lower PAE indicates stronger simplification). To remove redundant levels, we retain a matched set of distinct simplification levels per dataset based on unique PAE values, ensuring comparable granularity across techniques.

%% file: sec.study_design.tex
\section{Study Design}
We recruited 30 participants from Prolific~\cite{palan2018prolific} for our IRB-approved user study. Participants ranged in age from 18 to 65+. The sample was balanced by gender (15 men, 14 women, 1 preferred not to say). Most participants held at least a bachelor's degree (97\%), and visualization experience varied: 14 identified as intermediate users, 8 as beginners, 5 as advanced, and 3 as novices. Participants were compensated at Prolific’s standard rate.

The study was conducted via a web-based interface on a desktop or laptop. Participants were encouraged to think aloud, and with consent, we recorded audio and screen activity. They configured charts simultaneously for three target device sizes: tablet (1536×2048px), mobile (750×1334px), and smartwatch (324×394px), consistent with prior responsive visualization research~\cite{setlur2021semantic, blumenstein2016evaluating}. Across all devices, charts were displayed in landscape orientation. The tablet chart measured 1000px wide and 375px high; the mobile and smartwatch versions preserved the same aspect ratio through proportional scaling. In C2 and C3, participants could explore different simplification techniques within each device preview to compare how outputs appeared at that specific screen size. After adjusting the simplification level and, in C3, optionally selecting important points, participants finalized their chosen configuration before proceeding.

\subsection{Design and Conditions}
We employed a within-subjects design in which each participant configured all nine datasets. We evaluated three levels of simplification control: C1 (single pre-assigned technique with level control), C2 (six techniques with level control), and C3 (C2 with added manual point selection). These conditions isolate the effects of technique-specific constraint, technique choices, and mixed-initiative refinement. Dataset–condition assignments followed a 3×3 Latin square design. Three assignment patterns rotated conditions across datasets, and participants were evenly distributed across patterns. Tasks were ordered in a round-robin sequence so that no two consecutive scenarios shared the same dataset or condition, reducing learning effects. In C1, the simplification technique was randomized. Each participant encountered three C1 scenarios, each using a different technique selected from the six available options, with no repetition within a session. Exposure was approximately balanced across participants, ensuring that differences between C1 and C2/C3 reflect choice availability rather than specific algorithm behavior.

\begin{figure}[!t]
    \centering
    \captionsetup[subfloat]{labelfont=scriptsize,textfont=scriptsize}

    \includegraphics[width=0.4\linewidth, trim=10 0 10 5, clip]{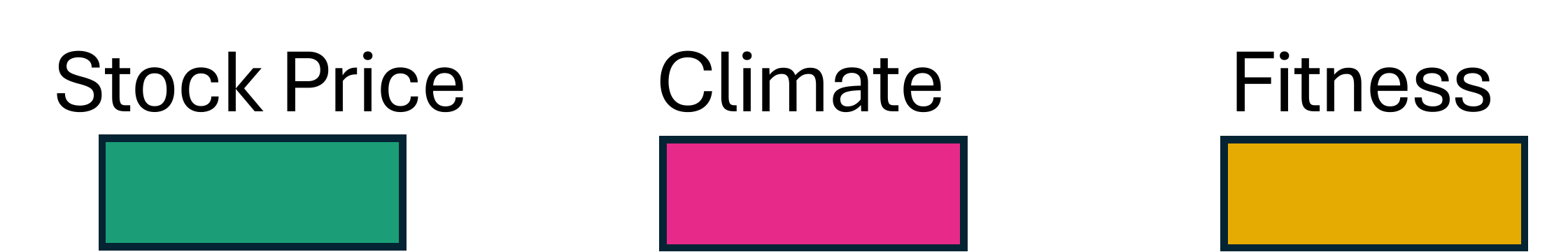}

    \vspace{6pt}

    \subfloat{
        \includegraphics[width=0.48\linewidth]{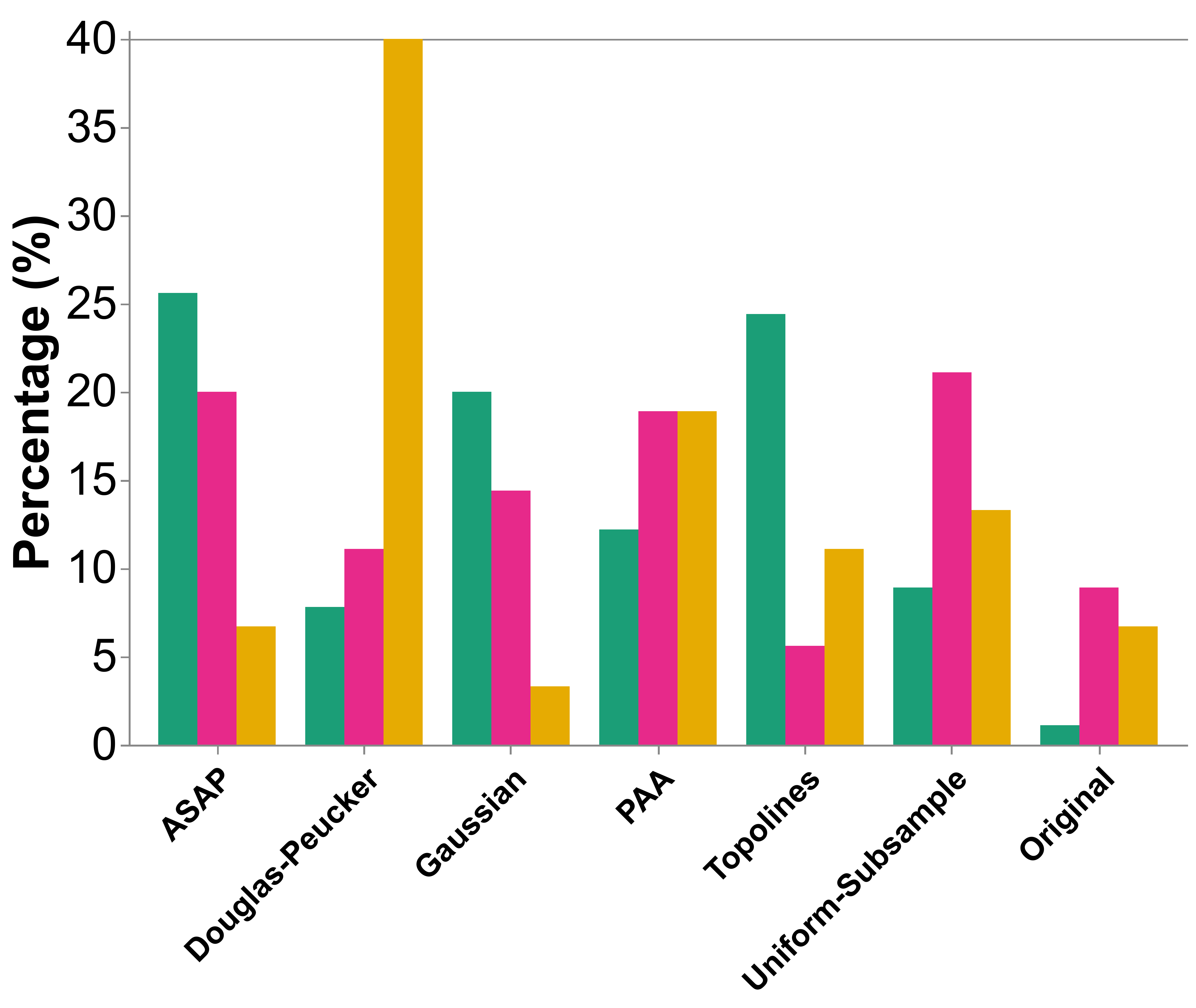}
    }
    \hfill
    \subfloat{
        \includegraphics[width=0.48\linewidth]{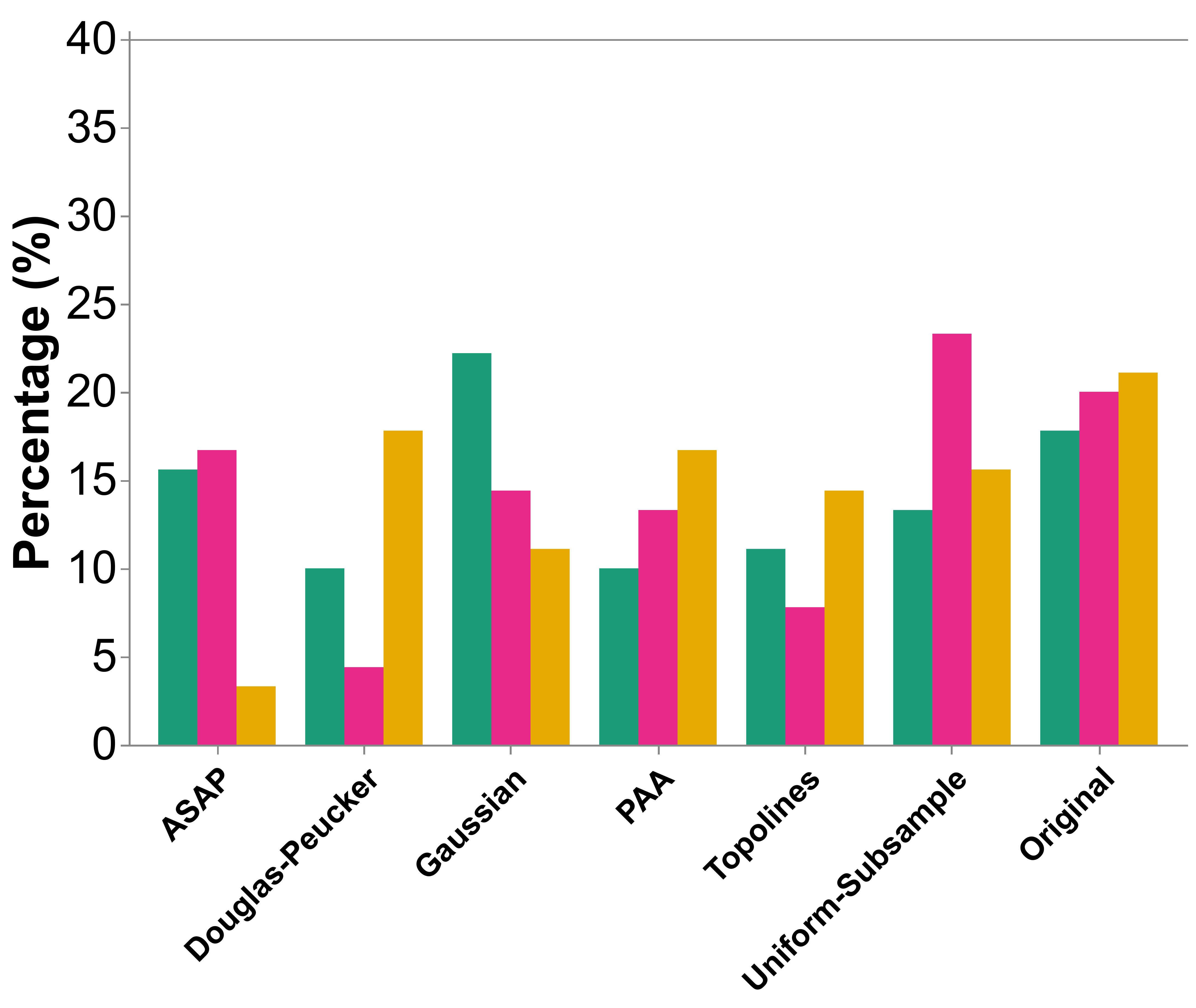}
    }

    \caption{Final simplification technique distribution across dataset categories in C2 (left) and C3 (right). Bars represent the percentage of finalized configurations using each technique, grouped by dataset type (Stock, Climate, Fitness).}
    \label{fig:algorithm_usage_c2_c3}
\end{figure}

\vspace{-10pt}

\subsection{Procedure and Measures}
Participants first completed a short tutorial using an unrelated practice dataset. They then completed nine configuration tasks. For each dataset, participants received a brief scenario describing the visualization context. Company names, wind-speed locations, and algorithm names for simplification were anonymized to reduce bias. All three device previews were displayed simultaneously, and participants independently configured simplification settings for each device with real-time feedback. After completing all nine tasks, participants completed a post-study questionnaire that captured overall preferences, reflections on strategies, and the condition (C1/C2/C3) they found most helpful. In addition to audio and screencast recordings, we logged algorithm selections, simplification levels, number of algorithm switches, task completion time, and manual point selections (C3 only).

%% file: sec.results.tex
\section{Results}
We analyzed 810 device configurations (30 participants × 9 datasets × 3 devices) distributed across three interface conditions (270 configurations per condition). Through interaction logs, post-study questionnaires, and think-aloud recordings, we characterized how participants approached simplification under varying levels of control. Our analysis yielded the following findings.

\begin{figure}[!t]
    \centering
    \captionsetup[subfloat]{labelfont=scriptsize,textfont=scriptsize}

    \subfloat[Participant preference\label{fig:helpful}]{
        \includegraphics[width=0.31\linewidth]{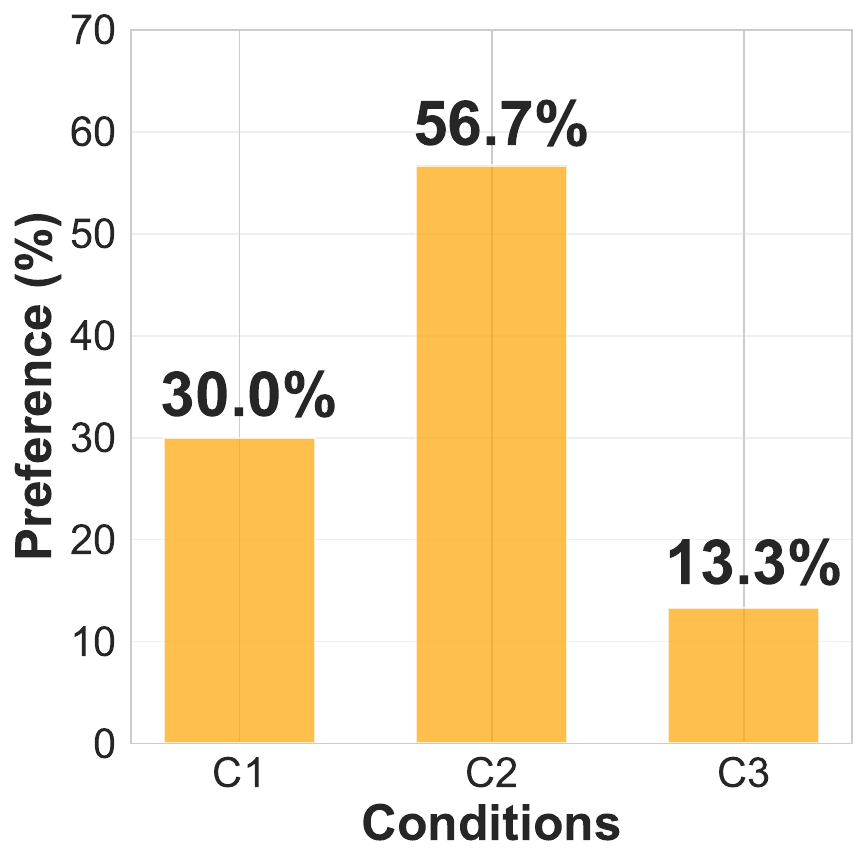}
    }
    \hfill
    \subfloat[Slider adjustments\label{fig:slider}]{
        \includegraphics[width=0.31\linewidth]{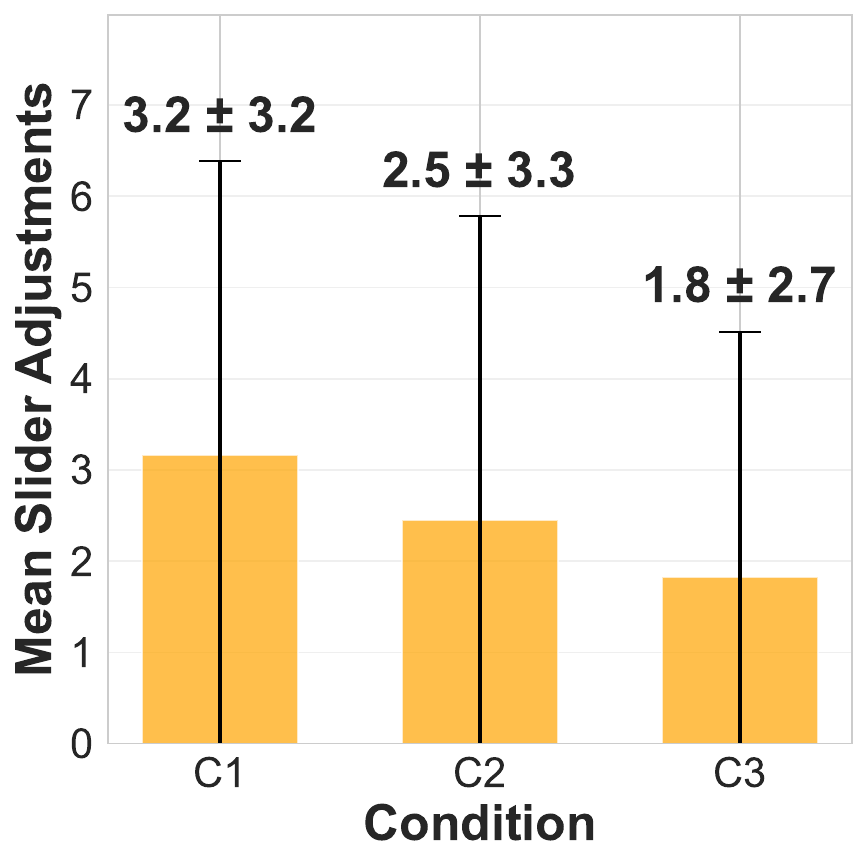}
    }
    \hfill
    \subfloat[Point selection usage\label{fig:points}]{
        \includegraphics[width=0.31\linewidth]{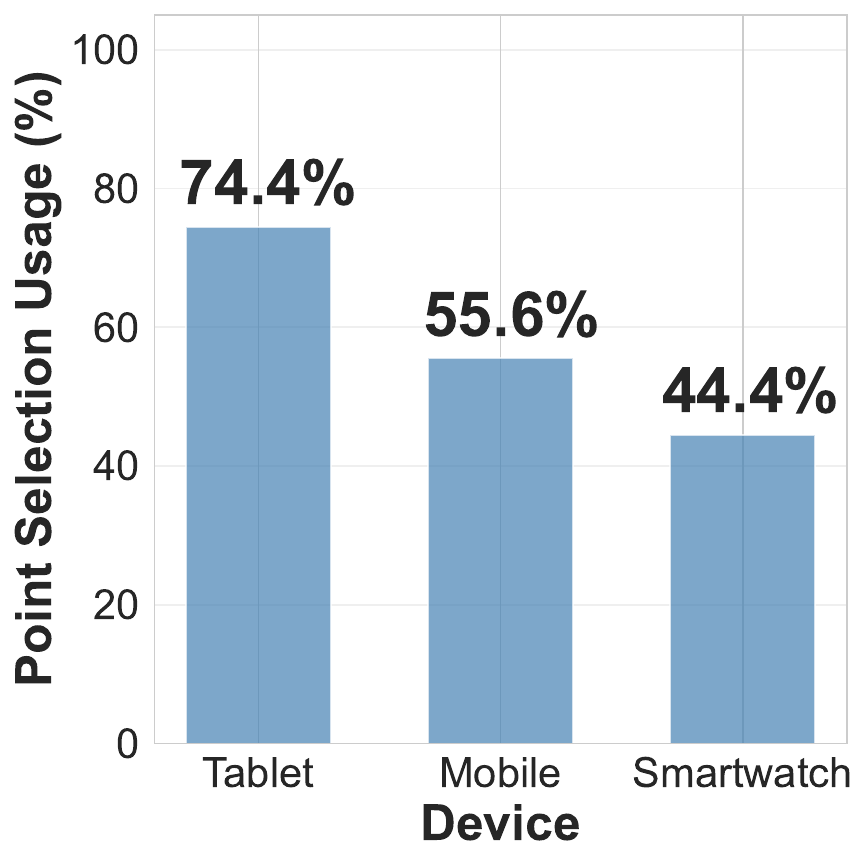}
    }

    \caption{Exploration and adaptation behavior. (a) Preferred condition (n=30). (b) Mean slider adjustments (n=270 per condition); error bars show standard deviation (c) Point selection usage by device in C3 (n=90 per device).}
    \label{fig:exploration_behavior}
\end{figure}
\vspace{-10pt}

\subsection{Interface Preferences: Diverse but Not Universal}
Post-study questionnaires revealed varied preferences across conditions (see~\Cref{fig:helpful}). While the majority preferred multiple techniques (C2: 56.7\%), 30\% favored the simplicity of a single technique (C1) and 13.3\% chose manual point selection (C3). This distribution challenged the assumptions about a universal best interface, suggesting that different participants optimized for different goals.

Qualitative responses reinforced this split. C2 supporters emphasized control. Participant \emph{P17} mentioned \emph{``I could compare multiple starting lines and find one that best matched the important points''}. C1 supporters valued efficiency. \emph{P3} mentioned \emph{``I felt more focused and decisive... reduced my hesitation and second-guessing''}. C3's low preference stemmed from perceived tediousness despite acknowledged utility for preserving critical features.

\subsection{Context-Driven Exploration and Technique Selection}
Although participants preferred having multiple techniques available, exploration within datasets was limited. Across C2 and C3, participants explored 2.66 unique techniques (out of 6 available) on average per dataset (across 3 devices), indicating minimal within-dataset exploration. With technique names anonymized and position randomized, this consistency suggests participants identified and selected techniques based on visual similarity across 
devices. However, across the full study, participants explored an average of 5.60 unique techniques, suggesting that choice was leveraged across datasets.

Technique preferences exhibited some dataset category-specific patterns (\Cref{fig:algorithm_usage_c2_c3}). For stock data in C2, \methodASAP{} (25.6\%) and \methodTDA{} (24.4\%), both preserving salient peaks, were most common. Fitness data strongly favored \methodDP{} (40.0\%), while climate data showed somewhat balanced usage. However, the broad distribution within each category (no technique exceeded 26\% for stock or climate) revealed individualized selection: different participants chose different techniques for similar datasets within the same category. These patterns shifted in C3, with a greater tendency to retain the ``original'' data for fitness, suggesting that manual point selection altered the strategy for selecting simplification techniques.

\subsection{Adaptation Across Control Dimensions}
Participants adapted their exploration strategy based on available control dimensions. In C1, where technique choice was constrained, participants made more slider adjustments (mean=3.2$, SD=$3.2) compared to C2 (mean=2.5$, SD=$3.3) and C3 (mean=1.8$, SD=$2.7) (see~\Cref{fig:slider}). We observed that participants compensated by exploring whichever control dimension was available: when technique selection was unavailable, they invested more effort in refining the simplification level. Manual point selection (C3) was used in 58.1\% of scenarios where available, with variation by device: tablet (74.4\%), mobile (55.6\%), smartwatch (44.4\%) (see~\Cref{fig:points}). When used, participants selected an average of 6.99 points, suggesting focused preservation of critical features. However, C3 also exhibited the highest rate of unsimplified configurations, indicating that added control complexity led some participants to disengage entirely.

%% file: sec.discussion.tex
\section{Discussion \& Conclusion}

Our findings suggest that the value of algorithmic choice depends on where it is introduced. Participants rarely switched techniques within a dataset but used multiple techniques across the study, indicating that their choices supported dataset-level strategy selection rather than repeated device-level optimization. Technique preferences shifted by data type (\methodASAP{} for volatile stock data, \methodDP{} for sparse fitness data), after which simplification levels were tuned to accommodate device constraints. The 56.7\% who preferred multiple techniques appeared to value this cross-dataset adaptability.

When technique choice was restricted (C1), participants compensated by exploring simplification levels more extensively. At the same time, 30\% preferred the constrained interface, revealing meaningful diversity in workflow preferences: some users prioritized decisiveness and reduced cognitive overhead over flexibility. Although manual point selection (C3) was used in 58.1\% of scenarios, it also resulted in the highest proportion of unsimplified configurations, suggesting that added interaction complexity did not consistently increase engagement.

\textbf{Design Implications.} The diversity of technique choices for similar datasets indicates that ``optimal'' simplification is subjective rather than universal. Responsive simplification systems should prioritize dataset-level technique selection while minimizing device-level reconfiguration burden. Rather than prescribing single ``best'' techniques, dataset-aware recommendations with explicit trade-off information (e.g., peak preservation vs. noise reduction) can support informed choices, while strong defaults preserve efficiency.

We studied preference over performance because the open-ended task lacks ground truth. Moreover, time-bounded sessions with crowd workers, rather than professional designers, may have amplified efficiency-oriented strategies. Within these constraints, algorithmic choice was most effective when enabling cross-dataset adaptation, and less impactful when repeatedly required at the device level. In future work, we plan to evaluate whether interfaces that incorporate these design recommendations improve designer workflows and enhance viewer comprehension across devices.

%% file: sec.acknowledgement.tex
\section*{Acknowledgments}
This work was supported by a grant from the National Science Foundation (III-2316496).